\documentclass[final,5p,times,twocolumn]{elsarticle}
\usepackage{graphicx} 
                       
\usepackage{epstopdf}
\graphicspath{{figures/}{figures/Wchannel/elecs/}}
\usepackage{atlasphysics} 
\usepackage{subfigure}
\usepackage{verbatim}
\usepackage{multirow}
\usepackage{lineno}
\usepackage{hyperref}

\def \MET { E_{T}^{\rm miss} }

\graphicspath {{figures/}{figures/Wchannel/elecs/}}

\begin{document}
\begin{frontmatter}

\title{\vspace{-0.5cm} \flushright{\normalsize{CERN-PH-EP-2011-193}} \\ \center{Search for heavy vector-like quarks coupling to light quarks in proton-proton \\ collisions at $\sqrt{s}=7$~TeV with the ATLAS detector}}

\author{The ATLAS Collaboration}

\date{\today}

\journal{Physics Letters B}

\begin{abstract}
This letter presents a search for singly produced vector-like quarks,
$Q$, coupling to light quarks, $q$. The search is sensitive to both
charged current (CC) and neutral current (NC) processes, $p p \to Q q
\to W q q' $ and $p p \to Q q \to Z q q' $ with a leptonic decay of
the vector gauge boson. In $1.04$~fb$^{-1}$ of data taken in 2011 by
the ATLAS experiment at a center-of-mass energy $\sqrt{s}=7$~TeV, no
evidence of such heavy vector-like quarks is observed above the
expected Standard Model background.  Limits on the heavy vector-like
quark production cross section times branching ratio as a function of
mass $m_Q$ are obtained.  For a coupling $\kappa_{qQ}= v/m_Q$, where
$v$ is the Higgs vacuum expectation value, 95\% C.L. lower limits on
the mass of a vector-like quark are set at 900~GeV and 760~GeV from CC
and NC processes, respectively.
\end{abstract}
\end{frontmatter}

\section{Introduction}
\label{sec:introduction}
Vector-like quarks (VLQ),
defined as quarks for which both chiralities have the same
transformation properties under the electroweak group $SU(2)\times U(1)$, 
are predicted by many extensions of the 
SM, relating to Grand Unification, dynamical electroweak symmetry breaking scenarios
or theories with extra dimensions~\cite{Hewett:1988xc,Andre:2003wc,Schmaltz:2005ky,Anastasiou:2009rv,Dobrescu:1997nm,Hill:2002ap,He:1999vp,He:2001fz,Choudhury:2001hs,Appelquist:2000nn}.
Since the couplings of the light quarks are well constrained, if VLQs 
exist they are generally expected to only couple sizably to the third generation.
However, in certain scenarios, corrections to quark mixings can
cancel, relaxing these constraints. 
The motivation and phenomenology of  heavy VLQs coupling to light generations is
discussed for the Tevatron~\cite{Atre:2008iu}, where a
baseline model is introduced which considers two
degenerate VLQ doublets having hypercharges 1/6 and 7/6 and mixing
only with the up quark. This scenario can occur naturally in certain models~\cite{Agashe:2006at}.
Because the doublets are degenerate, cancellations occur which allow VLQ coupling to the
first two generations, leading to a potentially strong signal at
the Large Hadron Collider (LHC).  

Following the notation of more recent work~\cite{Atre:2011ae} which
describes a model-independent approach to VLQ sensitivity at the LHC,
a coupling $\kappa_{qQ}=(v/m_Q) \tilde{\kappa}_{qQ}$ is defined here,
where $q$ stands for any light quark, $Q$ is the VLQ, $m_Q$ is the
VLQ mass, $v$ is the Higgs vacuum expectation value and
$\tilde{\kappa}_{qQ}$ encodes all the model dependence of the $qVQ$
vertex ($V=W$ or $Z$). 
Electroweak precision measurements constrain the contribution of heavy quarks
to loop diagrams, but under certain conditions, as for the degenerate VLQ doublet model above, 
mild bounds apply on the dimensionless coupling
$\kappa$, allowing it to be as large as $\sim 1$~\cite{Atre:2011ae}.
The masses of VLQs are not constrained by vacuum stability in the SM~\cite{Frampton:1999xi}.

\begin{figure}[h]
\centering
\includegraphics[width=0.45\textwidth]{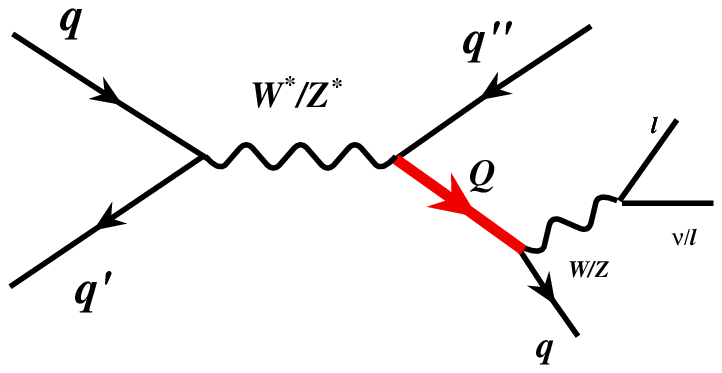}
\hspace{1cm}
\includegraphics[width=0.38\textwidth]{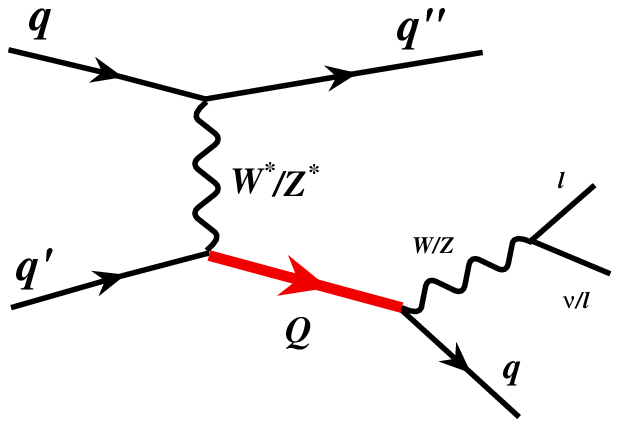}
\caption{Vector-like quark production and decay diagrams for
  $s$-channel (top) and $t$-channel (bottom). The thick line indicates the vector-like quark.}
\label{fig:diagram}
\end{figure}

It has been shown that single production
provides a favorable process to probe for the existence of these heavy
quarks if the coupling to light quarks is large, and that a significant
mass reach could be achieved at the LHC with early data~\cite{Atre:2008iu,Atre:2011ae}.  Single
production of a VLQ occurs via the process $ q q' \stackrel
{V^*} {\longrightarrow} q'' Q $ 
(Fig.~\ref{fig:diagram}).  A quark produced by this process
of gauge boson exchange can have a charge of $5/3,
2/3, -1/3$ or $ -4/3$. As a benchmark, we consider theories
with only VLQs $U$ of charge $+2/3$ or only with $D$ of charge $-1/3$,
without regard to the multiplet structure of the model.
The experimental limits obtained on cross section times branching ratio can
then be interpreted as limits on the couplings for different VLQ models~\cite{Atre:2011ae}.
The contribution from the
$s$-channel diagram is negligible compared to that of the $t$-channel
process. Therefore one characteristic of the signal is the presence of
a forward jet: after one of the initial state quarks emits the
electroweak gauge boson, it will continue in the forward direction with
little transverse momentum ($p_T$), while the other quark couples to the $W$ or $Z$ to
produce the heavy quark. Because the LHC is a proton-proton collider, the 
charged current (CC) production of a $D$ quark is expected to have a 
higher cross section than that of a $U$ quark. Similarly, for the 
neutral current (NC) process, $U$ quarks are expected to be produced more 
abundantly.  Anti-quark production is suppressed since it 
involves antiquarks in the initial state.

Bounds on the mass of new heavy quarks were obtained previously from a search in
the pair production process at the
Tevatron~\cite{Aaltonen:2009nr,Abazov:2011vy} and LHC~\cite{ATLAS-CONF-2011-022,Chatrchyan:2011ay}.
Limits have also been obtained at the Tevatron~\cite{cdf_hq,
  Abazov:2010ku} on single production processes $\sigma(p \bar p \to
qQ) \times BR(Q \to qW)$, which in the model~\cite{Atre:2008iu}
of degenerate doublets with $\tilde{\kappa}_{uD} = 1$ and decaying
100\% via CC gives a 95\% confidence level (C.L.) upper limit exclusion for $D$ quarks with masses up to 690~GeV. 
Limits at 95\% C.L. on $\sigma(p \bar p \to qQ)\times BR(Q\to qZ)$ in the
same model yield an exclusion of a $U$ quark with
$\tilde{\kappa}_{uU} = \sqrt{2}$ and 100\% branching ratio via NC up
to a mass of 550 GeV.

This Letter reports on a search for singly produced VLQs in the
ATLAS detector at the LHC. 
The search is conducted in events with at least two jets and a vector boson, indicated by either two high-$p_T$ leptons (electrons or muons) in the case of a $Z$, or a single lepton and missing transverse momentum ($\MET$) in the case of a $W$.
The data used in this analysis were collected from March to June 2011,
at a center of mass energy $\sqrt{s}= 7$~TeV and correspond to an integrated
luminosity of $\mathcal{L}= (1.04 \pm 0.04)$ fb$^{-1}$~\cite{ATLAS-CONF-2011-116}.

\section{The ATLAS Detector}
\label{sec:detector}
The ATLAS detector is a multi-purpose particle physics detector system
optimized to record information coming from $pp$
collisions~\cite{Aad:2008zzm}.  Closest to the interaction point is
the inner detector (ID) for charged particle tracking, which is
performed by silicon pixel and microstrip detectors in addition to a 
straw-tube tracker with radiators to produce transition radiation. The 
tracking system is embedded in a 2~T axial magnetic field. Surrounding 
the solenoid are the lead and liquid argon electromagnetic (EM) 
calorimeter and hadronic tile calorimeter subsystems. Forward 
calorimetry is accomplished with liquid argon detectors and copper 
and tungsten absorbers. These systems allow the reconstruction of
electrons and jets, both essential for this analysis.
Surrounding the calorimeter systems is a muon spectrometer (MS) that 
uses drift chambers to record muon trajectories in a toroidal magnetic 
field.  A three-level trigger is used to select events for subsequent 
offline analysis. Events recorded when a subsystem was not properly functioning
are not used in this analysis.

\section{Signal and background modeling}
\label{sec:sim}
Signal Monte Carlo (MC) samples are generated using
MadGraph~\cite{Maltoni:MadGraph} based on
Ref.~\cite{Atre:2008iu,Atre:2011ae}, then hadronized and showered
through {\tt PYTHIA}~\cite{Sjostrand:2006za}. The CTEQ6L1 parton
distribution function (PDF)~\cite{Pumplin:2002vw} is used, with
factorization and renormalization scales of $m_W$ ($m_Z$) for the CC
(NC) channel. Nine reference masses are generated for both CC and NC
decays: 225~GeV, 300~GeV, then continuing in steps of 100~GeV up to
1~TeV. The production cross section times branching ratio to a vector
boson and jets ranges from 194~pb to 0.47~pb for CC and from 88~pb to
0.28~pb for NC, assuming $\tilde{\kappa}_{qQ} = 1$.

The dominant SM backgrounds are $W \to \ell \nu$ + jets and $Z \to
\ell \ell$ + jets for the CC and NC channels, respectively. Other
sources of background are from multijet events, $t\bar t$, single top,
and diboson processes, which can have electrons or muons and jets in
the final states. With the exception of multijets, the contributions
of these backgrounds are estimated using MC samples.
$W$ + jets and $Z$ + jets samples are generated by {\tt
  ALPGEN}~\cite{Mangano:2002ea} using CTEQ6L1 PDFs with parton
showering performed by {\tt HERWIG}~\cite{Corcella:2000bw} and using
{\tt JIMMY}~\cite{Butterworth:1996zw} for simulation of the underlying
event model.  The cross section times leptonic branching ratios are
10.3~pb and 1.06~pb per lepton flavor for $W$ and $Z$'s, respectively,
with $p_T$ of the leptons~$>$~20~GeV. This includes $K$-factors of
1.22 and 1.25, respectively, to reproduce the inclusive cross sections
at next-to-leading order in QCD~\cite{Melnikov:2006kv}.
{\tt MC@NLO}~\cite{Frixione:2002ik} is used to
simulate $t\bar{t}$ production, giving a cross section of 165~pb. 
Single top quark events decaying leptonically ($\sigma = 37.5$~pb) 
are generated with {\tt
AcerMC}~\cite{Kersevan:2004yg} combined with parton showering and
hadronization by {\tt PYTHIA}.
Diboson backgrounds are simulated with {\tt ALPGEN} and {\tt HERWIG}
parton shower for the NC channel ($\sigma \times BR = 5.97$~pb), which
requires two leptons in the final state, and standalone {\tt HERWIG}
(with a $K$-factor of 1.52 to reproduce the inclusive cross section at
next-to-leading order in QCD~\cite{Campbell:2011bn}) for the CC
channel ($\sigma \times BR = 69.1$~pb) where a single lepton is
required.
Multijet backgrounds from QCD processes are derived both from {\tt
  PYTHIA} and data samples, described below.

The detector response simulation~\cite{:2010wqa} is based on {\tt
  GEANT4}~\cite{Agostinelli:2002hh,Allison:2006ve}.  The MC samples
are generated with superimposed minimum bias events to simulate the
conditions that occur in data. 
In order to improve the modeling of both signal and backgrounds, 
lepton reconstruction and identification efficiencies, energy scales
and resolutions in the MC are corrected to correspond to the values
measured in the data.

\section{Analysis}
\label{sec:analysis}
The analysis is subdivided into four channels: charged and neutral
current, each with either electrons or muons in the final state. 
Particle definitions and selections are identical in all channels, 
but signal and control regions for the CC and NC channels are defined independently.

Events are selected in which there is at least one
vertex reconstructed with at least three tracks. The vertex with the
greatest total transverse momentum, $\sum{|p_T|}$, of the
associated tracks is designated as the primary vertex. The trigger
requires at least one cluster in the EM calorimeter with $p_T >
20$~GeV or at least one muon candidate in the MS with a track
originating from the primary vertex with $p_T > 18$~GeV. In both
cases, the trigger requires a matching ID track.

Electron candidates are required to pass tight quality selection criteria based on
the calorimeter shower shape, track quality and track matching with
the calorimeter cluster~\cite{Aad:1390775}. They must have $p_{T} > 25$~GeV and lie in the
pseudorapidity\footnote{ATLAS uses a right-handed coordinate system with the $z$-axis along the beam pipe. 
               The $x$-axis points to the center of the LHC ring, and the $y$-axis points upward. 
               The pseudorapidity is defined in terms of the polar angle 
               $\theta$ as $\eta = -\ln \tan(\theta/2)$.} 
region $|\eta| < 2.47$, excluding
the regions of transition between the central and forward detector
sub-elements, $1.37 < |\eta| < 1.52$.
During most run periods of the data set, a region of the EM
calorimeter corresponding to about 1\% of channels was less efficient
than the rest of the detector. An exclusion window around the affected
area was defined as $-0.1 < \eta < +1.5$ in pseudorapidity and $-0.9 <
\phi < -0.5$ in azimuth. Electrons in this region are removed from
data collected during these periods. The same procedure is applied to
simulated events corresponding to the fraction of data covered by
these run periods. Finally, no more than 4~GeV of transverse energy is
allowed outside the core of the electron defined by a cone of size
$\Delta R=\sqrt{(\Delta\eta)^2+(\Delta\phi)^2}=0.2$.

Muon candidates are reconstructed by combining tracks from both the
ID and the MS.  They are required to pass ID quality requirements~\cite{ATLAS-CONF-2011-063} and have 
$p_{T} > 25$ GeV and $|\eta| < 2.4$. To suppress cosmic rays,
muon candidates must have a distance of closest approach to the
primary vertex in the longitudinal direction $|z_0| < 5$~mm and in the
transverse plane $|d_0| < 0.1$~mm.
Isolated muons are selected by requiring that the sum of ID track transverse momentum around the muon track, in
a cone of $\Delta R = 0.2$ divided by the $p_{T}$ of the
muon itself be less than 0.1.

Jet four-vectors are reconstructed from calorimeter clusters using the
anti-k$_T$ algorithm~\cite{Cacciari:2008gp} with a radius parameter of
0.4. After correcting for calorimeter
non-compensation and inhomogeneities by using $p_T$ and $\eta$
dependent calibration factors~\cite{ATLAS-CONF-2011-032}, jets are required to have $p_T >
25$~GeV and $|\eta| < 4.5$.
Events containing jets that fail quality
criteria~\cite{ATLAS-CONF-2010-038} are rejected to ensure an accurate
$\MET$ measurement.
Furthermore, events containing jets passing through the inefficient 
region of the EM calorimeter are vetoed.
To remove jets originating from other $pp$ interactions within an event, the selected jets are required to have more than 75\% of $p_T$-weighted ID tracks associated to the primary vertex.
Finally, to avoid counting electrons as jets, any jet candidate 
within $\Delta R < 0.2$ of a selected electron is removed.

The $\MET$ is calculated as the negative vector of the transverse
components of energy deposits in the calorimeters within $|\eta| <
4.5$. For events containing muons, any calorimeter energy deposit from
a muon is ignored and the muon energy measured in the MS is used
instead~\cite{Aad:1379858}.

The CC candidates are required to have (i)
exactly one electron or muon,
(ii) missing transverse momentum $\MET > 50$~GeV,
(iii) one jet with $p_T>50$~GeV and at least one more jet with
$p_T>25$~GeV, (iv) a minimum pseudorapidity separation $|\Delta \eta|
> 1.0$ between the highest-$p_T$ (leading) jet and second or
third-leading jet, since the presence of a forward jet is expected in
signal events, (v) $m_{T}(\ell,\MET) > 40$~GeV, where
$m_{T}(\ell,\MET) = \sqrt{2 E_T^{\ell} \MET (1 - \cos{\Delta
    \phi_{\ell,\MET}})}$ is the transverse mass of the $W$ candidate,
and (vi) an azimuthal angle separation between the lepton and $\MET$
vector $\Delta \phi_{\ell,\MET} < 2.4$~rad since the $W$ in the signal
is expected to be boosted.  To reconstruct the mass of the VLQ
candidate, the longitudinal momentum $p_{z}$ of the neutrino is
calculated such that the invariant mass of the lepton and $\MET$
equals the mass of the $W$. Of the two solutions, the one which leads
to the larger value of $|\Delta \eta|$ between the reconstructed
neutrino four-vector and the leading jet is chosen, since the
simulation shows it to be the correct solution about 60\% of the
time. If no real solution is found, the real part of the complex
solutions is taken. The system composed of the leading jet and the
reconstructed $W$ is taken to be the VLQ candidate.

The NC candidates are required to have exactly two oppositely
charged same-flavor leptons with an invariant mass in the range $66 <
M(\ell,\ell) < 116$~GeV and a transverse momentum
$p_{T}(\ell,\ell) > 50$~GeV. At least two jets of $p_T > 25$~GeV are
required, with the same $|\Delta \eta| > 1.0$ requirement as
described for the CC selection. The invariant mass of the system
composed of the two leptons and the leading jet is taken to be the
VLQ candidate mass.

To evaluate the level of multijet background in the CC analysis, a
procedure is used based on a fit to the $\MET$ distribution in the
range $0 < \MET < 100$~GeV. For this purpose, only selection criteria
(i) and (v) above are required.
For both the electron and muon modes, template shapes for the
non-multijet backgrounds are taken from the MC samples described
earlier and summed according to their relative cross sections. The
overall normalization of this non-multijet template is left
floating. In the electron mode, a sample enriched in objects
misidentified as electrons (fakes) is selected from data using {\it
  medium} quality electrons, excluding {\it tight} electrons, as
defined in~\cite{Aad:1390775}. The $\MET$ distribution of this sample
serves as the electron multijet template shape. For the muon mode,
multijet background is primarily expected to come from heavy flavor
decays. Therefore, the multijet template shape is taken from a {\tt
  PYTHIA} sample of $b\overline{b}$ events. For both lepton flavors, a
fit to the $\MET$ distribution is performed using the multijet and
non-multijet templates to determine the normalization of the multijet
component.
The modeling of the multijet background was tested in a control region
defined by the range of $10 < \MET < 30$~GeV. The modeling of the
kinematic variable distributions, and in particular of the VLQ
candidate mass, was found to agree with data within statistics.

In the case of the NC selection, the multijet background is estimated from data-driven studies
to be negligible. In the electron channel the selection for electron
candidates is changed to require the medium criteria, excluding tight
electrons, to obtain a fake di-electron template shape in $m_{ee}$,
which is then scaled to make the total background expectation match
the data with the same selection. In the muon channel the isolation
requirement is inverted, and the $m_{\mu\mu}$ template scaled to the 
data in the same way.

\begin{figure}
  \centering
  \subfigure{ 
 	\includegraphics[width=0.49\textwidth] {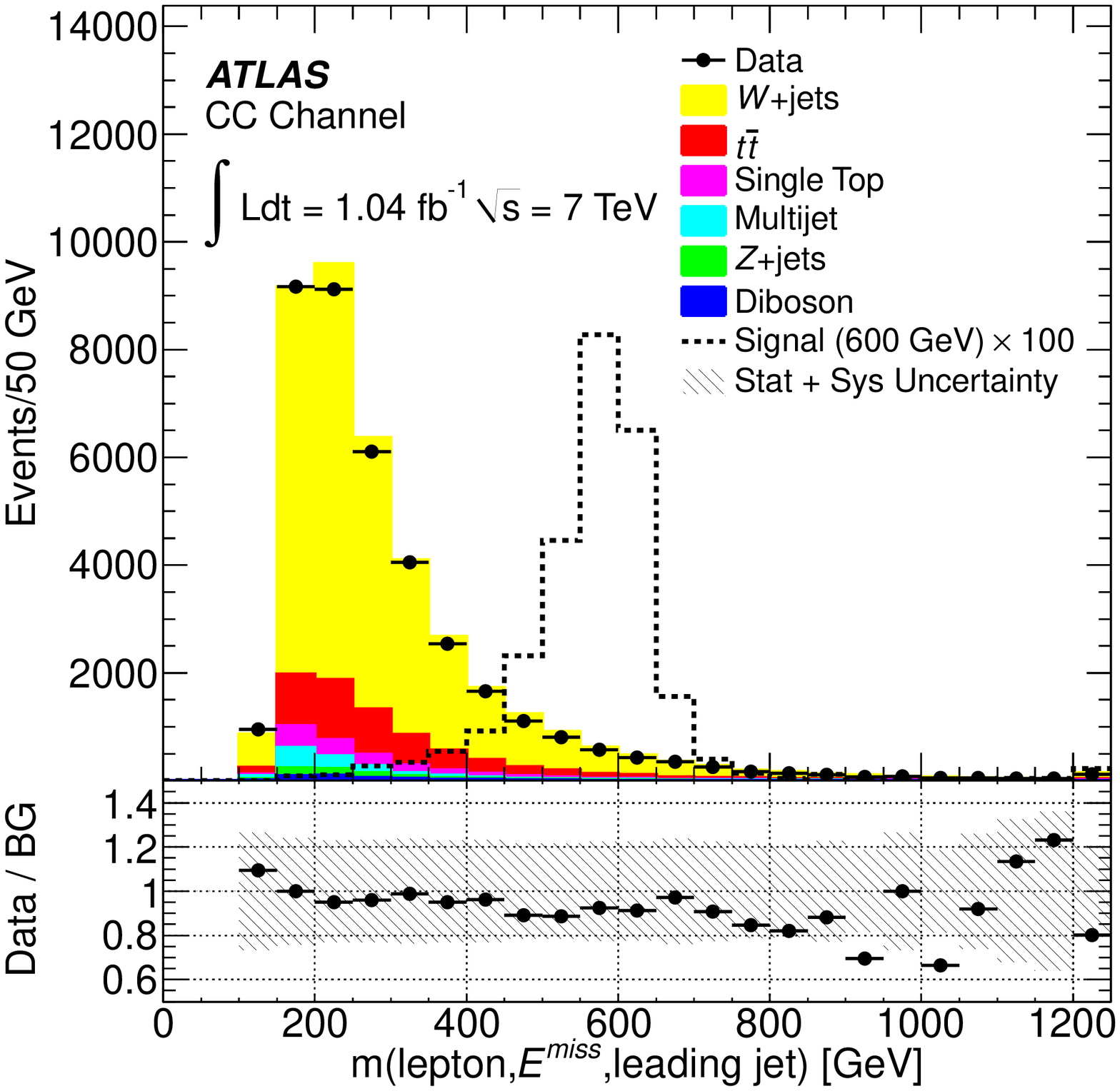}
    \label{fig:W_M}
  }
  \subfigure{
	 \includegraphics[width=0.49\textwidth] {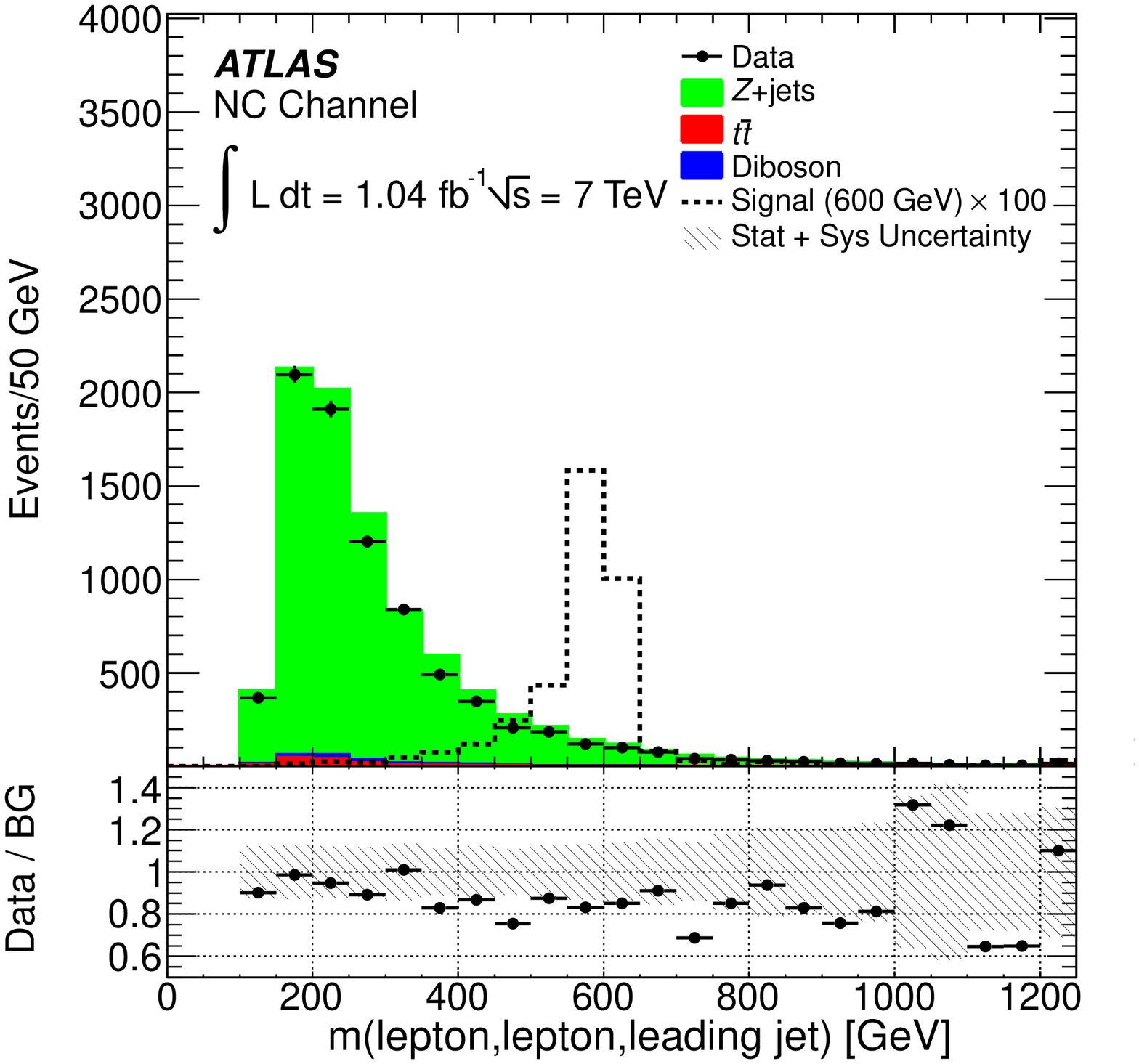}
    \label{fig:Z_M}
  }
  \caption{Invariant mass distribution of VLQ candidates in both the
    CC (top) and NC (bottom) channels, summed over both the electron and
    muon final states. The dashed line shows the signal shape,
    normalized by 100 times the leading order theoretical cross
    section. The bottom part of each plot shows the ratio of the data
    to the background model. The last bin contains events with
    invariant mass candidates equal to or higher than 1200 GeV.}
  \label{fig:invmass}
\end{figure}

With the above selections, the observed event yields and corresponding
predictions are given in Tables~\ref{tab:w_SR_eventcounts}
and~\ref{tab:nc_cutflow_elec}.  
From these yields, no significant excess is observed in the data that
can be accounted for by a VLQ signal.  Figure~\ref{fig:invmass} shows
the invariant mass distributions of the reconstructed VLQ candidate in
the signal regions for both channels. These distributions are used in
a binned likelihood fit to extract signal yields and production cross
section upper limits. A slight shape discrepancy between data and MC
is apparent in Figure~\ref{fig:invmass}. Before extracting an upper
limit, a correction is applied to the MC background shape, as
described in Section~\ref{sec:results}.

\begin{table*}[htb]
 \begin{center}
   \begin{tabular}{|c|c|c|} \hline
    Process                                  &       Electron channel & Muon channel \\ \hline
    
    \hline
    $W$+jets                                 &  $ 14500 \pm 100 \pm 4400 $     &  $ 16600 \pm 100 \pm 5000 $    \\ \hline
    $t\bar{t}$                               &  $  2360 \pm 50 \pm 270 $       &  $  2530 \pm 50 \pm 290 $        \\ \hline
    Single Top                               &  $   700 \pm 30 \pm 120 $       &  $   740 \pm 27 \pm 120 $        \\ \hline
    Multijet                                 &  $   670 \pm 30 \pm 270 $       &  $   340 \pm 20 \pm 410 $       \\ \hline
    $Z$+jets                                 &  $   128 \pm 11 \pm 90 $        &  $   432 \pm 21 \pm 170 $        \\ \hline
    Diboson                                  &  $   174 \pm 13 \pm 53 $        &  $   198 \pm 14 \pm 62 $         \\ \hline
    \hline
    Expected Total Background                &  $ 18500 \pm 100 \pm 4400 $     &  $ 20900 \pm 100 \pm 5100 $    \\ \hline 
    Data                                     &  $ 17302 $                      &  $ 20668 $                     \\ \hline
    \hline
    Expected Signal, $D$(225 GeV)            &  $ 2360 \pm 50 \pm 350 $        &  $ 2380 \pm 50 \pm 400 $      \\ \hline
    Expected Signal, $D$(600 GeV)            &  $ 133 \pm 12 \pm 10 $          &  $ 133 \pm 12 \pm 11 $      \\ \hline
    Expected Signal, $D$(1000 GeV)           &  $ 14  \pm 4    \pm 1 $       &  $ 14 \pm 4 \pm 1 $    \\ \hline
  \end{tabular}
  \caption{Expected and observed event counts in the kinematically
    allowed VLQ mass range after the final selection in the CC channel
    with an integrated luminosity of 1.04~$\ifb$.
Uncertainties are statistical and systematic, respectively. 
The signal predictions assume a coupling $\tilde{\kappa}_{uD}=1$. }

  \label{tab:w_SR_eventcounts}
 \end{center}
\end{table*}

\begin{table*}[htb]
\begin{center}
\begin{tabular}{|c|c|c|}
\hline
       Process            & Electron Channel & Muon Channel \\ \hline
       \hline
       $Z$+jets           & 3250 $\pm$ 60 $\pm$ 430     & 5350 $\pm$ 70 $\pm$ 700   \\ \hline
       $t\bar{t}$         & 58 $\pm$ 8 $\pm$ 3      & 90 $\pm$ 9 $\pm$ 5     \\ \hline
       Diboson            & 38 $\pm$ 6 $\pm$ 4     & 58 $\pm$ 8 $\pm$ 4     \\ \hline
       \hline
       Expected Total Background  & 3350 $\pm$ 60 $\pm$ 430  & 5500 $\pm$ 70 $\pm$ 700   \\ \hline
       Data                   & 3105  & 5070  \\ \hline
       \hline
       Expected Signal, U(225 GeV)         & 192 $\pm$ 14 $\pm$ 9   & 339 $\pm$ 18 $\pm$ 19   \\ \hline
       Expected Signal, U(600 GeV)         & 15 $\pm$ 3.9 $\pm$ 0.6    & 23 $\pm$ 4.8 $\pm$ 0.7   \\ \hline
       Expected Signal, U(1000 GeV)        & 1.9 $\pm$ 1.4 $\pm$ 0.1   & 2.7 $\pm$ 1.6 $\pm$ 0.1   \\ \hline
\end{tabular}
  \caption{Expected and observed event counts in the kinematically
    allowed VLQ mass range after the final selection in the NC channel
    with an integrated luminosity of 1.04~$\ifb$. Uncertainties are
    statistical and systematic, respectively.
The signal predictions assume a coupling $\tilde{\kappa}_{uU}=1$.}
\label{tab:nc_cutflow_elec}
\end{center}
\end{table*}

\section{Systematic uncertainties}
\label{sec:systematics}

Systematic uncertainties on the simulation of the signal arise from
uncertainties in PDFs and the factorization and renormalization scales. In
order to estimate the uncertainty due to the parton distributions,
the CTEQ66~\cite{Nadolsky:2008zw} PDF set is used, for which the
eigenvectors of the Hessian matrix are known.
The difference in signal cross section due to the PDF uncertainty is
found to range from 3.0\% at a signal mass of 225~GeV to 4.4\% at
1000~GeV.  The uncertainty due to the factorization and
renormalization scales is estimated by taking the difference between
signal cross sections at the nominal value of the scales, and at
values of one-half and twice the nominal. The uncertainty is found to
vary between 4\% and 12\% for the same mass range. Uncertainties due
to the simulation of initial and final state radiation are found to be
about 1\%. These uncertainties on the theoretical cross section are added in quadrature.

For signal and background events, the jet-energy-scale uncertainty is
calculated by shifting the $p_{T}$ of all jets up and down by factors
that vary as a function of $p_{T}$ and $\eta$. The factors range from
4.6\% for jets with $p_{T} = 20$~GeV to 2.5\% for jets with $p_{T}$
above 60~GeV~\cite{ATLAS-CONF-2011-032}. This procedure results in an
uncertainty of about 20\% on the background normalization, and about
5\% on the signal efficiency.
The jet-energy-resolution uncertainty is calculated by smearing the
$p_{T}$ of each jet depending on the jet
$p_{T}$ and $\eta$, typically by around 10\%. This
source of uncertainty is found to impact both the background
normalization and signal efficiency by about 1\%.
The lepton-energy-scale uncertainty is evaluated and found to be much
less than 1\% for both signal and background. The effect of the
previously mentioned EM calorimeter inefficiency is also found to be
much less than 1\%. 
Uncertainties also arise from the trigger, identification, and
reconstruction efficiency corrections applied to the MC
simulation. They affect the signal efficiency uncertainty by 1-2\%
depending on the mass.
The rate uncertainty from MC statistics after event selection is 3-5\%.
Finally, the uncertainty on the luminosity is
3.7\%~\cite{ATLAS-CONF-2011-116}. None of the systematics studied have
been found to significantly affect the shape of the VLQ candidate mass
distribution.

\section{Results}
\label{sec:results}

To determine signal yields, a binned maximum likelihood fit is
performed using template histograms of the VLQ candidate mass
distribution. The fit is performed separately for each signal
mass. The electron and muon final states are fitted
simultaneously. The overall signal and background normalizations are
left floating in the fit. Systematic uncertainties on the template
normalizations are incorporated as Gaussian-distributed nuisance
parameters, as are the signal efficiency systematics used in
determining the cross section limits.
Signal template shapes are taken from MC, while background templates
are as shown in Fig.~\ref{fig:invmass}, with an additional correction
described next.

A heavy VLQ signal would appear as a peak on top of a smooth
background in the VLQ candidate invariant mass distribution.
It is therefore important to have a good estimate of the background
shape in the region around a signal mass hypothesis. The fit procedure
described above makes use of the full range of mass, but the
normalization is dominated by the lower mass region where the number
of events is higher. A small shape difference between Monte Carlo and
data can therefore yield a systematic bias in the fit at high
mass. For that reason a correction is applied to the background model
for each signal mass. It is obtained from linear fits to the
reconstructed invariant mass of the ratio of data/MC after the full
event selection, excluding bins in the range [-200,+100]~GeV around
each signal mass tested.  The asymmetric choice in the excluded mass
is motivated by the fact that the expected signal has a low mass tail.
The 1$\sigma$ uncertainty in the slope is taken as a systematic shape
uncertainty.  It was verified that no significant difference to the
fit results arose from choosing a narrower excluded mass window, or
even no exclusion at all.

Since no significant excess of data over the background prediction is
observed in either channel, limits as function of the VLQ mass are
obtained based on the likelihood fits. Pseudoexperiments are generated
by sampling the likelihood function to compute the expected limits,
using a Gaussian prior for all nuisance parameters and including the
shape uncertainty from the linear correction.

\begin{figure}
  \centering
  \subfigure{ \includegraphics[width=0.49\textwidth]{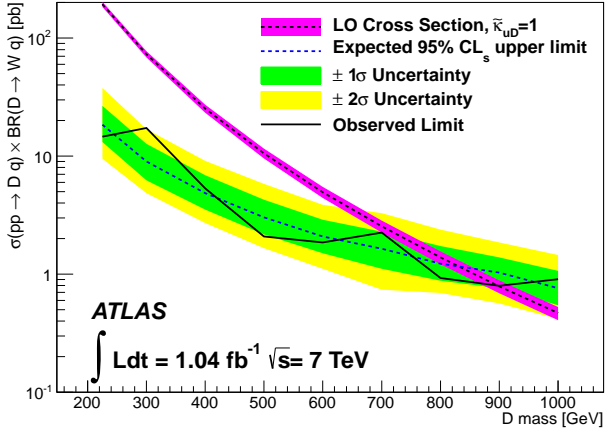} }
  \subfigure{ \includegraphics[width=0.49\textwidth]{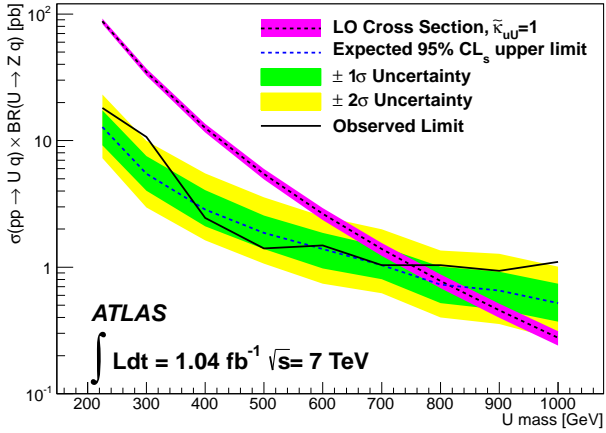} } 
  \caption{Upper limits at 95\% confidence level on the cross section
    times branching ratio $\sigma(pp\to Qq) \times BR(Q\to Vq)$ for
    the CC (top) and NC (bottom) channels as a function of mass. The
    leading-order (LO) theoretical cross section assumes
    $\tilde{\kappa}_{uD}=1$ and $\tilde{\kappa}_{uU}=1$ on the top and
    bottom respectively. The width of the dark band around it
    corresponds to the theoretical uncertainty described in the
    text. The expected cross section upper limit is determined by the
    median result of background-only pseudoexperiments, and is shown
    with its $1 \sigma$ and $2\sigma$ uncertainties, respectively.}
  \label{fig:cls_limits}
\end{figure}

The 95\% C.L. exclusion limits on $\sigma(p p \to Q q) \times BR(Q\to
Vq)$ as a function of the VLQ mass, based on the CL$_s$
method~\cite{Read:2002}, are shown in
Fig.~\ref{fig:cls_limits}. Taking the intersection of the observed
(expected) cross section limits with the central value of the
theoretical cross section, masses below 900~GeV (840~GeV) are excluded
for the CC channel and 760~GeV (820~GeV) for the NC channel, assuming
a coupling $\tilde{\kappa}_{qQ}^2=1$ and a 100\% branching ratio for
VLQs to decay to a vector boson and a jet. Within the $\pm 1\sigma$
theoretical uncertainties, the observed CC mass limit ranges from $870
- 920$~GeV. The corresponding range for the NC limit is $730 -
770$~GeV.
Limits for each mass tested are given in Table~\ref{tab:limitResults}. The
fourth and fifth columns show an interpretation of the cross section
limits in terms of limits on the couplings $\tilde{\kappa}_{uD}^2$ and
$\tilde{\kappa}_{uU}^2$, in each case assuming only $D$ production or
only $U$ production, respectively, and 100\% branching fraction to a
vector boson and jet.

A stronger limit in the CC channel may be
obtained by repeating the CC analysis, requiring a negatively charged lepton because the SM
background from $W^-$+jets is lower than for $W^+$+jets. The upper
limits on $\sigma(p p \to D^{-\frac{1}{3}} q) \times BR(D^{-\frac{1}{3}}
\to W^- u)$ are given in the sixth column of
Table~\ref{tab:limitResults}.

\begin{table*}[htb]
\begin{center}
  \begin{tabular}{|c|c|c|c|c|c|} \hline
    Mass [GeV] & CC $\sigma \times$ BR [pb] & NC $\sigma \times$ BR [pb] & $\tilde{\kappa}_{uD}^2$ & $\tilde{\kappa}_{uU}^2$ & CC$^-$ $\sigma \times$ BR [pb]  \\ \hline
    225        & 15                               & 18                               & 0.075                        & 0.21  & 12                          \\ \hline
    300        & 17                               & 11                               & 0.24                         & 0.31  & 5.6                          \\ \hline
    400        & 5.3                              & 2.4                              & 0.21                         & 0.19  & 3.8                         \\ \hline
    500        & 2.1                              & 1.4                              & 0.19                         & 0.26  & 1.1                         \\ \hline
    600        & 1.9                              & 1.5                              & 0.37                         & 0.56  & 1.9                         \\ \hline
    700        & 2.2                              & 1.0                              & 0.86                         & 0.75  & 2.2                         \\ \hline
    800        & 0.93                             & 1.0                              & 0.66                         & 1.33  & 0.97                        \\ \hline
    900        & 0.80                             & 0.9                              & 1.0                          & 2.1   & 0.70                        \\ \hline
    1000       & 0.91                             & 1.1                              & 1.9                          & 4.0   & 0.50                        \\ \hline
  \end{tabular}
 \caption{Observed upper limits at 95\% confidence level on the cross
   section times branching ratio $\sigma(pp\to Qq) \times BR(Q\to V
   q)$ as a function of mass and the corresponding upper limit on a
   model-independent heavy-to-light quark coupling. The final column
   shows the limit on the CC process after selecting
   negatively charged leptons.}
 \label{tab:limitResults}
\end{center}
\end{table*}

\section{Conclusion}
\label{conclusion}

A search for single production of vector-like quarks coupling to light
generations has been presented. No evidence is found for such quarks
above the expected background in either the CC or NC channel.  Upper
limits on the production cross section times branching ratio to a
vector boson and a jet were determined at 95\% confidence
level. Assuming couplings $\tilde{\kappa}_{uD}^2 =1$ and
$\tilde{\kappa}_{uU}^2 =1$, the upper bounds obtained for the mass of
vector-like quarks are 900~GeV for the CC channel and 760~GeV for the
NC channel. These limits, which can be used to constrain different
models of vector-like quarks~\cite{Atre:2011ae}, are the most
stringent to date on this benchmark model.

\section{Acknowledgments}
\label{sec:acknowledgments}

We thank A.~Atre, M.~Carena, T.~Han, and J.~Santiago for the MadGraph
code used to produce the signal MC samples.

We thank CERN for the very successful operation of the LHC, as well as the
support staff from our institutions without whom ATLAS could not be
operated efficiently.

We acknowledge the support of ANPCyT, Argentina; YerPhI, Armenia; ARC,
Australia; BMWF, Austria; ANAS, Azerbaijan; SSTC, Belarus; CNPq and FAPESP,
Brazil; NSERC, NRC and CFI, Canada; CERN; CONICYT, Chile; CAS, MOST and
NSFC, China; COLCIENCIAS, Colombia; MSMT CR, MPO CR and VSC CR, Czech
Republic; DNRF, DNSRC and Lundbeck Foundation, Denmark; ARTEMIS, European
Union; IN2P3-CNRS, CEA-DSM/IRFU, France; GNAS, Georgia; BMBF, DFG, HGF, MPG
and AvH Foundation, Germany; GSRT, Greece; ISF, MINERVA, GIF, DIP and
Benoziyo Center, Israel; INFN, Italy; MEXT and JSPS, Japan; CNRST, Morocco;
FOM and NWO, Netherlands; RCN, Norway; MNiSW, Poland; GRICES and FCT,
Portugal; MERYS (MECTS), Romania; MES of Russia and ROSATOM, Russian
Federation; JINR; MSTD, Serbia; MSSR, Slovakia; ARRS and MVZT, Slovenia;
DST/NRF, South Africa; MICINN, Spain; SRC and Wallenberg Foundation,
Sweden; SER, SNSF and Cantons of Bern and Geneva, Switzerland; NSC, Taiwan;
TAEK, Turkey; STFC, the Royal Society and Leverhulme Trust, United Kingdom;
DOE and NSF, United States of America.

The crucial computing support from all WLCG partners is acknowledged
gratefully, in particular from CERN and the ATLAS Tier-1 facilities at
TRIUMF (Canada), NDGF (Denmark, Norway, Sweden), CC-IN2P3 (France),
KIT/GridKA (Germany), INFN-CNAF (Italy), NL-T1 (Netherlands), PIC (Spain),
ASGC (Taiwan), RAL (UK) and BNL (USA) and in the Tier-2 facilities
worldwide.

\onecolumn
\begin{flushleft}
{\Large The ATLAS Collaboration}

\bigskip

G.~Aad$^{\rm 48}$,
B.~Abbott$^{\rm 110}$,
J.~Abdallah$^{\rm 11}$,
A.A.~Abdelalim$^{\rm 49}$,
A.~Abdesselam$^{\rm 117}$,
O.~Abdinov$^{\rm 10}$,
B.~Abi$^{\rm 111}$,
M.~Abolins$^{\rm 87}$,
H.~Abramowicz$^{\rm 152}$,
H.~Abreu$^{\rm 114}$
% ...
\bigskip

% ...
$^{10}$ Institute of Physics, Azerbaijan Academy of Sciences, Baku, Azerbaijan\\
$^{11}$ Institut de F\'isica d'Altes Energies and Departament de F\'isica de la Universitat Aut\`onoma  de Barcelona and ICREA, Barcelona, Spain\\
% ...
$^{48}$ Fakult\"{a}t f\"{u}r Mathematik und Physik, Albert-Ludwigs-Universit\"{a}t, Freiburg i.Br., Germany\\
$^{49}$ Section de Physique, Universit\'e de Gen\`eve, Geneva, Switzerland\\
% ...
$^{87}$ Department of Physics and Astronomy, Michigan State University, East Lansing MI, United States of America\\
% ...
$^{110}$ Homer L. Dodge Department of Physics and Astronomy, University of Oklahoma, Norman OK, United States of America\\
$^{111}$ Department of Physics, Oklahoma State University, Stillwater OK, United States of America\\
% ...
$^{114}$ LAL, Univ. Paris-Sud and CNRS/IN2P3, Orsay, France\\
% ...
$^{117}$ Department of Physics, Oxford University, Oxford, United Kingdom\\
% ...
$^{152}$ Raymond and Beverly Sackler School of Physics and Astronomy, Tel Aviv University, Tel Aviv, Israel\\
% ...
\end{flushleft}


\end{document}